\documentclass{article}

\begin{document}

\title{Motion of Poly-vectors in Fractal Clifford Spaces }         
\date{}          
\maketitle
\begin{center}
{\author {Magd E. Kahil{\footnote{ Faculty of Engineering, Modern Sciences and Arts University, Giza, Egypt  \\
e.mail: mkahil@msa.edu.eg}}} }
\end{center}
\abstract{
 Fractal Clifford Spaces (FCS) may be considered   as a challenging approach to the unification of micro-physics and macro-physics. Trajectories of these manifolds are described by different poly-vectors describing paths and their deviation equations for not only to test particles but also for charged, rotating and rotating charged objects. Due to relaxing the condition of differentiability for all extended objects in the manifold. In this approach, a specialized Bazanski Lagrangian will be discussed to describe the fractal version of extended objects in a fractal space-time. }

\section{Introduction}
The problem of unification has to be revisited every now and then, providing a new description of the structure of space-time such as replacing point like by p-branes as described within Clifford space, the building blocks of this manifolds is based on poly-vectors rather than vectors [1] and types of differentiability involving fractals [2]. This may lead many authors [3] to propose some aspects that play a vital role to present it.

Following this track, it is worth mentioning to examine motion of poly-vectors as defined in a fractal Clifford Space as an approach towards unification of micro and macro physics. This farfetched dream may require thinking outside the box in the realm of relaxing the usual common assumptions that keep each field to be treated independently. In our study we are going to describe a fractal space time which is continuous but not differentiable. This step may give rise to widen the scope of work to expect not only one geodesic between two points but infinite geodesics [4]. This would lead to consider two different types of differentiation, forward and backward. Such an ensemble makes to describe velocity in a complex space [5].

The paper is of concerned with revisiting the notation of path and path deviation equations fractal and their counterparts of spinning and spinning deviation objects in a specific type of Clifford space describing gravity using fractal derivatives in order to observe their associated manifold such is continuous and not differentiable.

\subsection{Fractal Space-times: A brief Introduction}
 Trajectories of nature are fractal, the necessity, we study curves that are continuous but not differentiable, and this may give rise to define two types of differentiations forward and backward. This makes the number of geodesics infinite between two arbitrary points in that prescribed space. A reliable approach to understand nature is based on systems that are continuous but not differentiable. One can demonstrate that the length of continuous and nowhere differentiable curve is dependent on resolution $\epsilon \rightarrow 0$ and further that ${\it{L}}(\epsilon) \rightarrow \infty$ when $\epsilon \rightarrow 0$, the curve is fractal (Nottale 1996).  Scale relativity theory (SRT) is an approach to revisit quantum mechanics from different domains of scale laws ranging from microscopic to macroscopic.
 If the particle is subject to the Brownian motion of unknown origin which is described by two Winer Processes (forward - backward), when they combine together they yield the complex nature of the wave function and they perform Newton's equation of motion into the Schroedinger equation [5,6].

Nature needs to be revisited from different perspectives, due to sudden discrepancies to understand the cause of striking problems in cosmology like dark energy and in dark . This may led many scientists to become skeptic in all methods of computation and detection.
Yet, one of good approaches to start revisiting the laws of nature from their building blocks.  In other words,  relaxing conditions that are fundamentally crucial in their constructions. This may led us to derive equations of motion for objects described not only by points but also as p-branes, in a space not differentiable but fractal. So, examining the behavior of particles spinning, charged or even both in fractal space-time, whose trajectories are continuous but not differentiable.

Such a confinement may be in need to introduce its underlying mechanism which enable us to describe them. We also express the corresponding Bazanski method which may be represented in this realm.
\subsection{ Fractal Geometry: An overview}
 Agop et al [7]that  point out  in Notalle's model of scale relativity, it is supposed that micro particles are taking place on fractal curves with fractal dimensions $D_{F}$ defining a manifold of fractal space time. This one has the following characteristic behavior:  The number of geodesics is infinite between two arbitrary points; a statistical (fluid-like) description was accepted. \\ Each geodesic is considered a fractal curve [7]. \\  The breaking of differential time reflection invariance. The case on non-differentiability gives rise to define the following functions:
\begin{equation}
\frac{df_{+}}{dt} = \lim_{\Delta t \rightarrow 0^{+}} \frac{f(t+\Delta t,\Delta t )- f(t, \Delta t)}{\Delta t}
\end{equation}
and

\begin{equation}
\frac{df_{-}}{dt} = \lim_{\Delta t \rightarrow 0^{-}}\frac{f(t, \Delta t)-f(t- \Delta t, \Delta t)}{\Delta t}
\end{equation}
 According to scale relativity it implies to the following relations:
\begin{equation}
d X_{+} = v_{+}dt + d\xi_{+},
\end{equation}
and
\begin{equation}
d X_{-} = v_{-}dt + d\xi_{-}
\end{equation}
where $v_{+}$ and $v_{-}$ are forward and backward mean velocities such that

\begin{equation}
v_{+}=\frac{d X_{+}}{dt}= \lim_{\Delta t \rightarrow 0^{+}} < \frac{X(t +\Delta t)- X(t)}{\Delta t} >,
\end{equation}

and
\begin{equation}
v_{-}=\frac{d X_{-}}{dt}= \lim_{\Delta t \rightarrow 0^{-}} < \frac{X(t) -\Delta (t- \Delta t)}{\Delta t} >,
\end{equation}
 such as $d\xi_{\pm}$ is the measure of non-differentiability with properties
\begin{equation}
<d \xi_{\pm} > =0.
\end{equation}
If $\frac{(v_{+} + v_{-})}{2}$ is consider a differentiable speed , then $\frac{(v_{+} - v_{-})}{2}$ is counted to be a non-differentiable speed, such that $d_{\pm} V = d V_{\pm}$. \\
Agop et al (2006) described the associated fractal covariant derivative in the following way []:
\begin{equation}
\bar{\Gamma}^{\lambda}_{\mu \nu} = \frac{1}{2}\tilde{g}^{\lambda \gamma}(g_{\nu \gamma, \mu} +\tilde{g}_{\gamma \mu, \nu} - g_{\mu \nu, \gamma } ) + \frac{1}{2}\tilde{g}^{\lambda \gamma} (C^{\rho}_{\mu \nu}\tilde{g}_{\rho \gamma} + C^{\rho}_{\gamma \nu}\tilde{g}_{\rho \nu} - C^{\rho}_{\nu \gamma}\tilde{g}_{\rho \mu})
\end{equation}
where $ \bar{\Gamma}^{\lambda}_{\mu \nu}$ is the affine connection, $g_{\mu \nu}$ the metric tensor, and $C^{\lambda}_{\mu \nu}$ the structure  constants defined by the non-commutativity of the partial derivative $\bar{\partial}_{\rho}$ i.e.

\begin{equation}
 [\bar{\partial}_{\mu}, \bar{\partial}_{\nu}] = C^{\lambda}_{\mu \nu}\partial_{\lambda}
\end{equation}

Thus, the fractal covariant derivative of a vector $X^{\mu}$ is defined as follows:

\begin{equation}
\bar{\nabla}_{\rho} X^{\mu} = \bar{\partial}_{\rho} X^{\mu} + \bar{\Gamma}^{\mu}_{\delta \rho} X^{\delta}.
\end{equation}

Thus, the commutation of two covariant derivatives becomes
\begin{equation}
 [\bar{\nabla}_{\mu}, \bar{\nabla}_{\nu}] = C^{\lambda}_{\mu \nu}\bar{\nabla}_{\lambda},
\end{equation}

Also, the Riemannian curvature tensor $\bar{R}^{\sigma}_{\gamma \mu \nu}$ is defined as follows:
\begin{equation}
\bar{R}^{\rho}_{\gamma \mu \nu} = \bar{\Gamma}^{\rho}_{\gamma \nu, \mu} - \bar{\Gamma}^{\rho}_{\gamma \mu, \nu} +  \bar{\Gamma}^{\delta}_{\gamma \nu,}\bar{\Gamma}^{\rho}_{\delta \mu} - \bar{\Gamma}^{\delta}_{\gamma \mu,}\bar{\Gamma}^{\rho}_{\delta \nu} - C^{\delta}_{\mu \nu} \bar{\Gamma}^{\rho}_{\delta \gamma}
\end{equation}
The extra term $C^{\alpha}_{\mu \nu}\bar{\Gamma}^{\rho}_{\alpha \delta}$ is due to the effect of non-commutation  of the curvature tensor.

The torsion of fractal space time as anti-symmetric part of affine connection can be amended in a fractal space-time as defined [5]
\begin{equation}
\bar{\Lambda}^{A}_{BC} = \bar{\Gamma}^{A}_{BC} - \bar{\Gamma}^{A}_{CB} -\bar{C}^{A}_{BC}.
\end{equation}

Thus, it can be found that for an arbitrary poly-vector $A^{M}$,
\begin{equation}
\frac{\bar{\nabla} A^{M}}{\bar{\nabla} x^{\mu \nu}} = [\nabla_{\mu}, \nabla_{\nu}] A^{M} +  C^{N}_{\mu \nu}D_{N }A^{M}
\end{equation}
where , $\frac{\bar{\nabla}}{\bar{\nabla} x^{\mu \nu}}$ is the covariant derivative with respect to a plane $x^{\mu \nu}$,  such that

\begin{equation}
\frac{\bar{\nabla}s}{\bar{\nabla} x^{\mu \nu}} = [\bar{\nabla}_{\mu}, \bar{\nabla}_{\nu}] s= K^{\rho}_{\mu \nu} \partial_{\rho}s +  C^{\rho}_{\mu \nu}D_{\rho}s,
\end{equation}
where,  $K^{\rho}_{\mu \nu}$ is the torsion tensor.
\begin{equation}
\frac{\bar{\nabla} a^{\alpha}}{\bar{\nabla} x^{\mu \nu}} = [\bar{\nabla}_{\mu}, \bar{\nabla}_{\nu}] a^{\alpha}+ C^{\delta}_{\mu \nu}\bar{\nabla}_{\delta}a^{\alpha} =  \bar{R}^{\alpha}_{\rho \mu \nu} a^{\rho}+ K^{\rho}_{\mu \nu} \bar{\nabla}_{\rho}a^{\alpha} + C^{\rho}_{\mu \nu}\bar{\nabla}_{\rho}a^{\alpha} + C^{\delta}_{\mu \nu}\bar{\nabla}_{\delta}a^{\alpha}.
\end{equation}
Yet, this type of torsion (15) can related to the notation of torsion as mentioned by Hammond [8]  as commutation of the potential associated by a prescribed scalar field $\phi$ [8]
\begin{equation}
 K^{\alpha}_{\mu \nu} = \frac{1}{2}(\delta^{\alpha}_{\mu} \phi_{,\nu} -\delta^{\alpha}_{\nu} \phi_{,\mu} ) + C^{\delta} _{\mu \nu}{\phi_{,\delta}}.
\end{equation}
The fractal metric tensor is defined as [9]
\begin{equation}
\tilde{g}_{\mu \nu} = g_{\mu\nu} + \gamma_{\mu \nu}
\end{equation}
such that
$\gamma_{\mu \nu}$ is the fluctuation metric tensor (defined within fractal dimensions). \\
The fractal covariant derivative is defined as follows [3]:
\begin{equation}
\frac{\bar{\nabla}}{\bar{\nabla}S} =  \frac{{D}}{{D}S} + V^{\alpha}.{D}_{\alpha} - i \mu ({D}^{\mu}{D}_{\mu} + \xi R )
\end{equation}
where, $\frac{D}{DS}$ is the well-known covariant derivative as described in the context of Riemannian geometry.
Nowadays,  in which its corresponding geodesic may be described as follows:

\begin{equation}
\frac{\bar{\nabla}U^{\alpha}}{\bar{\nabla}S}=0.
\end{equation}
Also, the same technique can be applied to get the variation with respect to the tangent vector $U^{\rho}$ to get the geodesic deviation equations:
\begin{equation}
\frac{\bar{\nabla}^2\Psi^{\alpha}}{\bar{\nabla}S^{2}} = \bar{R}^{\alpha}_{. \beta \gamma \delta} \Psi^{\gamma} U^{\beta}U^{\delta}
\end{equation} The above method has been applied in different geometries than the Riemannian one e.g.
the above Lagrangian has been modified to describe the path equation of charged object to take the following form :
\begin{equation}
L = \tilde{g}_{\alpha \beta} \tilde{V}^{\alpha}{\frac{\tilde{D} \tilde{\Psi}^{\beta}}{\tilde{D}S}} + \tilde{f}_{\beta}\tilde{\Psi}^{\beta}
\end{equation}

where
$${ \tilde{f}_{\beta} =  a_{1} F_{\alpha \beta} \tilde{V}^{\beta} + a_{2} \bar{R}_{\alpha \beta \gamma \delta} \tilde{S}^{\gamma \delta} \tilde{V}^{\alpha}} ,$$

 $a_{1}$ and $a_{2}$ are parameters may take the following values ${\frac{e}{m}}$ and ${\frac{1}{2m}}$ to be adjusted with the original Lorentz force equation and Papapetrou equation [10] as well as Dixon equation [11].

\section{ Fractal Clifford Space: Underlying Geometry}
{{ In this section, we are going to introduce an extended object as described as a poly-vector existed  in a fractal space. A point in a Fractal  Clifford Space (FCS) is defined  as a set of holographic coordinates $(s,x^{\mu}, x^{\mu \nu},...)$ forming the coordinates of a poly-vector . Each one  is expressed within bases $\{\gamma_{A}\} =\{1, \gamma_{a_{1}},\gamma_{a_{1}a_{2}},\gamma_{a_{1}a_{2}a_{3}}....  \} $} , $a_{1}<a_{2}<a_{3}<a_{4}<a_{5}<...$, $r=1,2,3...., $ where $\gamma_{a_{1}a_{2}a_{3}...} = a_{1}\wedge a_{2}\wedge a_{3}....$  . \\ It is well known that the local basis $\gamma_{\mu}$, is related to the tetrad field $e^{a}_{\mu}$ such that [12] $$\gamma_{\mu} = e^{a}_{\mu}\gamma_{a} $$ in which
   \begin{equation}
   A= a + \frac{1}{1!}a^{\mu}\gamma_{\mu} + \frac{1}{2!}a^{\mu \nu}\gamma_{\mu} \wedge  \gamma_{\nu}+..........\frac{1}{n!}a^{\mu_{1}.....\mu_{n}} \gamma_{\mu_{1}} \wedge ......\gamma_{\mu_{n}}
   \end{equation}
The fractal derivative of C-space is defined as follows
\begin{equation}
\bar{d} A= \frac{\partial A}{\partial X^{B}}\bar{d}X^{B}
\end{equation}
i.e
\begin{equation}
\bar{d} A= \frac{\partial A}{\partial s} +  \frac{\partial A}{\partial x^{\mu}}\bar{d}x^{\mu}+  \frac{\partial A}{\partial x^{\mu \nu}}\bar{d}x^{\mu \nu}+......
\end{equation}
in other words, if one takes $A =\gamma_{\mu}$ , then
\begin{equation}
\bar{d} \gamma_{\mu}= \frac{\partial \gamma}{\partial s} +  \frac{\partial \gamma_{\mu}}{\partial x^{\mu}}\bar{d}x^{\mu}+  \frac{\partial \gamma_{\mu}}{\partial x^{\mu \nu}}\bar{d}x^{\mu \nu}+......
\end{equation}
which becomes\begin{equation}
\bar{d} \gamma_{\mu}= \frac{\partial \gamma}{\partial s} +  \Gamma^{\alpha}_{\mu \nu }\bar{d}x^{\nu}+  \Gamma^{\alpha}_{\mu [\nu , \rho] }\bar{d}x^{\nu \rho}+......
\end{equation}
which can be reduced to
\begin{equation}
\bar{d}\gamma_{\mu}= \frac{\partial \gamma}{\partial s} +
\bar{\Gamma}^{\alpha}_{\mu \nu }\bar{d}x^{\nu}+  \frac{1}{2}\bar{R}^{\alpha}_{\beta \nu \rho}\bar{d}x^{\nu \rho}+......
\end{equation}
where $\bar{R}^{\alpha}_{\beta \nu \rho}$ is the curvature of fractal space-time. \\
Also, it can be found that for an arbitrary poly-vector $A^{M}$
\begin{equation}
\frac{\bar{D} A^{M}}{\bar{D} x^{\mu \nu}} = [\bar{D}_{\mu}, \bar{D}_{\nu}] A^{M}
\end{equation}
where $\frac{\bar{D}}{\bar{D} x^{\mu \nu}}$ is the covariant derivative with respect to a plane $x^{\mu \nu}$,  such that

\begin{equation}
\frac{\bar{D} s}{\bar{D} x^{\mu \nu}} = [\bar{D}_{\mu}, \bar{D}_{\nu}] s= \bar{K}^{\rho}_{\mu \nu} \partial_{\rho}s,
\end{equation}
where $K^{\rho}_{\mu \nu}$ is the torsion tensor.
\begin{equation}
\frac{\bar{D} a^{\alpha}}{\bar{D} x^{\mu \nu}} = [\bar{D}_{\mu}, \bar{D}_{\nu}] a^{\alpha}=  \bar{R}^{\alpha}_{\rho \mu \nu} a^{\rho}+ \bar{K}^{\rho}_{\mu \nu} \bar{D}_{\rho}a^{\alpha},
\end{equation}
Yet, this type of torsion  can related to the notation of torsion as mentioned by Hammond [9] as commutator the potential associated by a prescribed scalar field $\phi$
\begin{equation}
 \bar{K}^{\alpha}_{\mu \nu} = \frac{1}{2}(\delta^{\alpha}_{\mu} \phi_{\bar{,}\nu} -\delta^{\alpha}_{\nu} \phi_{\bar{,}\mu} )
\end{equation}
 Thus, we can figure out that the torsion as defined in C-space (Riemannian Type) by means of the parameters $s$ may act like an independent scalar field defined in the usual context of Riemannian-Cartan geometry.

From examining equations (30) and (31), one can find the existence of torsion tensor in the even presence of symmetric affine connection apart from its conventional notation definition of being the anti-symmetric part of an affine connection as in the context of non-Riemannian geometries.  Accordingly, owing to C-space one may realize the dispute between the reliability of torsion propagating or non-propagating this can be resolved by means of of we describe torsion as a result of covariant differentiation of areas of holographic coordinates and non-propagating as being defined as anti-symmetric parts of an affine connection of poly-vectors or vectors, if one utilizes in the internal or external coordinate its corresponding non-symmetric affine connection.

Consequently, we can regard that the geometry described within the coordinates of poly-vector described not only Riemannian but also a non-Riemannian, i.e. the composition of a Riemannian affine connection for the poly-vector is not necessarily Riemannian as well, This may through some light to find out that a Riemannian Poly-vector affine connection and curvature (external coordinate  capital Latin letters) may be described by non- Riemannian quantities as describing its holographic coordinates (internal coordinate (Greek letters).

The arising notation of continuous and not spaces evolves the existence of a vital parameter called resolution [4]. This means that scale relativity is going to extending the original relativity. It is worth mentioning that Nottale 1993 initiated this concept paving the way to a huge number of papers examining different types of physics using scale relativity (SR). One of very fundamental aspects due to implementing notation of SR which gives rise that physical quantities are becoming scale-dependent [ref.]. In other words, the length $\it{L(\epsilon)}$ of the fractal curve is becoming function of resolution such that $$\it{L(\epsilon)}= \it{L_{0}(1+\frac{\lambda}{\epsilon})^{\delta}}$$
where $L_{0}$ the classical length  $\lambda$ is a constant to be fixed like the Planck length microscopic scale  or the Compton length for the macroscopic scale, $epsilon$ is scale of resolution acts as a fractal dimension: for better adjustment $\epsilon$ becomes $\ln{\epsilon}$ and $\delta$.
" Scale relativity is an approach to understand quantum mechanics, and physical domains to define chaotic system as well as large scale cosmological"
Accordingly the poly-vector may be expressed in fractal space as follows
\begin{equation}
X^{A} = X^{A}_{0}(1+ \frac{\lambda_{0}}{\Lambda} )^{\delta}
\end{equation}
Thus, in order to implement the concept of scale relativity on poly-vectors as defined in Clifford space, it has to allocate the quantities of  $\lambda$ and $\epsilon$ of scale relativity  to be specified as $\lambda_{0}$ the electron wave length and $\Lambda$ the Planck length.

Accordingly the poly-vector may be expressed in fractal space as follows
\begin{equation}
X^{A} = X^{A}_{0}(1+ \frac{\lambda_{0}}{\Lambda} )^{\delta}
\end{equation}

\section{The Bazanki Approach for Extended Objects of Fractal Clifford Space}
 The Bazanski approach [13] to obtain an equation of an object may be found as follows:
 \begin{equation}
 L= G_{AB} W^{A} \frac{\bar{D} \Phi^{B}}{\bar{D} \Xi}
 \end{equation}

 Taking the variation with respect to $ \Phi^{C}$ and $W^{C}$ we obtain after some manipulations one finds
 \begin{equation}
\frac{\bar{D}W^{C}}{\bar{D}S}=  0,
 \end{equation}

and
\begin{equation}
\frac{\bar{D}^2 \Phi^{C}}{\bar{D}\bar{S}^{2}}=  \bar{R}^{C}_{BDE} W^{E} W^{D} \Phi^{E} ,
 \end{equation}
where $\bar{;}$ is the fractal covariant derivative.

 \subsection{Spinning and Spinning Deviation Equations of a Fractal Clifford Space}
We suggest the equivalent Bazanski Lagrangian [14-19] for deriving the equations for spinning and spinning poly-vectors of a Fractal Clifford Space (FCS) to be
\begin{equation}
 L= G_{AB} P^{A} \frac{\bar{D} \Psi^{B}}{\bar{D}\bar{S}} + S_{AB} \frac{\bar{D} \Psi^{AB}}{\bar{D}\bar{S}}+ \bar{F}_{A}\Psi^{A}+ \bar{M}_{AB}\Psi^{AB}.
 \end{equation}
 such that $$ P^{A}= m U^{A}+ U_{\beta} \frac{\bar{D} S^{A B}}{\bar{D}S}$$
where $P^{\mu}$ is the momentum poly-vector $ \bar{F}^{\mu} = \frac{1}{2} \bar{R}^{\mu}_{\nu \rho \delta} S^{\rho \delta} U^{\nu}$, $\bar{R}^{\alpha}_{\beta \rho \sigma}$ is its corresponding  Riemann curvature, $\frac{\bar{D}}{\bar{D}\bar{S}}$ is the fractal  covariant derivative with respect  to a parameter $S$,$S^{\alpha \beta}$ is the spin poly-tensor, and $ M^{\mu \nu} =P^{\mu}U^{\nu}- P^{\nu}U^{\mu}$
 such that $U^{\alpha}= \frac{\bar{d} x^{\alpha}}{\bar{d}s}$ is the unit tangent poly-vector to the geodesic one.
 In a similar way as performed in (40), by taking the variation with respect to $ \Psi^{\mu}$ and$\Psi^{\mu \nu}$ simultaneously one obtains
 \begin{equation}
\frac{\bar{D}P^{M}}{\bar{D}\bar{S}}= \bar{F}^{M},
 \end{equation}
 \begin{equation}
\frac{\bar{D}S^{M N}}{\bar{D}\bar{S}}= \bar{M}^{M N} ,
 \end{equation}
  \\
 Using the following identity on both equations (41) and (42)
  \begin{equation}
  A^{D}_{\bar{;} N H} - A^{D}_{\bar{;} H N} = \bar{R}^{D}_{ B N H} A^{B},
  \end{equation}
  such that $A^{D}$ is an arbitrary poly-vector.
 Multiplying both sides with arbitrary poly-vectors, $U^{H} \Psi^{B}$ as well as using the following condition
 \begin{equation}
 U^{A}_{\bar{;} H} \Psi^{H} =  \Psi^{A}_{\bar{;} H } U^{H},
 \end{equation}
and $\Psi^{A}$ is its deviation poly-vector associated to the  unit poly-vector tangent $U^{A}$.
 Also in a similar way:

 \begin{equation}
 S^{AB}_{\bar{;} H} \Psi^{H} =  \Psi^{AB}_{\bar{;} H } U^{H},
\end{equation}

 We obtain the corresponding deviation equations:
  \begin{equation}
\frac{\bar{D}^{2}\Psi^{A}}{\bar{D}S^{2}}=  \bar{R}^{A}_{B H C}P^{B} U^{H} \Phi^{C}+ \bar{F}^{A}_{\bar{;} H} \Psi^{H},
 \end{equation}
and
 \begin{equation}
\frac{\bar{D}^{2}\Psi^{A B}}{\bar{D}\bar{S}^{2}}=  S^{[B D }{\bar{R}}^{A ]}_{ D H C} U^{H} \Psi^{C}+ \bar{M}^{AB}_{\bar{;} H} \Psi^{H}.
 \end{equation}
\subsection{The Generalized Dixon Equation in C-space }
As a byproduct, we suggest the following Lagrangian which enables us to obtain the spinning charged poly-vector in FCS .
\begin{equation}
 L= G_{AB} P^{A} \frac{\bar{D} \Psi^{B}}{\bar{D}S} + S_{AB} \frac{\bar{D} \Psi^{AB}}{\bar{D}S}+\tilde{F_{A}} \Psi^{A}U^{B}+ \tilde{M}_{AB} \Psi^{AB},
 \end{equation}
such that $$ \tilde{F_{A}}= \frac{1}{m}(e \bar{F}_{AB} + \frac{1}{2}{\bar{R}}_{ABCD}S^{CD})$$ and  $$\tilde{M}_{AB}= (P_{A}U^{B}-P_{B}U^{U} +  \bar{F}_{AC}S^{C}_{.~B}-\bar{F}_{AC}S_{A~.}^{.~C} )$$ .
 Taking the variation with respect to $ \Psi^{\mu}$ and$\Psi^{\mu \nu}$ simultaneously we obtain
\begin{equation}
\frac{\bar{D}P^{M}}{\bar{D}S}= \frac{\epsilon}{M} \bar{F}^{M}_{B}U^{B} + \frac{1}{2M} \bar{R}^{M}_{NEQ} S^{EQ}U^{N},
 \end{equation}
 \begin{equation}
\frac{\bar{D}S^{M N}}{\bar{D}S}= P^{M}U^{N}-P^{N}U^{M} + \bar{F}^{MC}S_{C}^{.~N} - \bar{F}^{CN}S^{M}_{.~C} .
 \end{equation}
Thus, applying the same laws of commutation as illustrated in (43)and(44), we obtain their corresponding deviation equation
 \begin{equation}
\frac{\bar{D}^{2}\Psi^{A}}{\bar{D}S^{2}}=  \bar{R}^{A}_{B H C}P^{B} U^{H} \Phi^{C}+ ( \frac{\epsilon}{M} \bar{F}^{M}_{B}U^{B} + \frac{1}{2m}{\bar{R}}^{M}_{NEQ} S^{EQ}U^{N})_{\bar{;} H} \Psi^{H},
 \end{equation}
and
 \begin{equation}
\frac{\bar{D}^{2}\Psi^{A B}}{\bar{D}S^{2}}=  S^{[B D }{R}^{A ]}_{ D H C} U^{H} \Psi^{C}+ (P^{M}U^{N}-P^{N}U^{M} + \bar{F}^{M C}S_{C}^{.~ N} - \bar{F}^{CN}S^{M}_{.~ C})_{\bar{;} H} \Psi^{H}.
 \end{equation}

\section{Conclusions}
In this paper we have defined a new type of path equations for extended objects as defined in a fractal space as an attempt to combine micro-physics with macro-physics. Owing to this idea one may express extended objects in Clifford space as points described in Fractal Space. Accordingly, their path and path deviation equations are obtained as in equation (30 and 31) not only those ones but their spinning and spinning deviation equations in as shown  in (46), (47), (49)and (50).\\
However, such an approach is regarded as a prototype to comprehend the unification problem. Yet, the work has to be extended using strong theories of gravity such as bi-metric theory in order to examine the problem of singularity as well as fluctuations of space time, which will be expressed in our future work.
\section*{References}
{[1]}C. Castro and M. Pavsic, Progress in Physics vol 1, 32 (2005) \\
{[2]} L. Nottale, Chaos, Solitons and Fractals vol 7 , No.6 ,877 (1996) \\
{[3]} M. E. Kahil and T. Harko , 85 Proceedings MEARIM- IAU Regional Meetings April 5-10,2008  , eds. A.A Hady and M.I. Wanas, Cairo, Egypt \\
{[4]} L. Nottale Astronomy and Astrophysics, {\bf{327}}, 867(1997) \\
{[5]} M. Agop and I. Gottieb, J. Math Phys. {\bf{47}}, 05353 (2006)\\
{[6]} D. da Rocha and L. Nottale Chaos, Solitons  and Fractals {\bf{16}}, 565, (2003). \\
{[7]} M. Agop and I. Gottieb, Eur. Phys. Journal D  {\bf{49}}, 239 (2008)\\
{[8]} R. Hammond {\it{Reports on Progress in physics}} {\bf{65}}, 599. \\
{[9]} L. Nottale, Marie- Noelle Celerier and T. Lerner arXiv: hep-th/0605280 (2006). \\
{[10]}A Papapetrou  {\it{Proceedings of Royal Society London A}} {\bf{209}} , 248(1951)  \\
{[11]} W.G. Dixon {\it{Proceedings of Royal Society London A}} {\bf{314}} , 499(190)  \\
{[12]} Magd E. Kahil Journal of Modern Physics (2020) 11,1865 DOI  : 10.4236/jmp.2020.1111116 .\\
{[13]} S.L. Bazanski  {\it{J. Math. Phys.}} {\bf {30}}, 1018 (1989) \\
{[14]}M.I. Wanas, M. Melek, and M.E. Kahil, Astrophys. Space Sci.,228, 273. ; gr-qc/0207113
(1995).\\
{[15]}M.I. Wanas, M. Melek, M.E. Kahil Gravit. Cosmol., {\bf{6} }, 319 (2000).  \\ 
{[16]}   Magd E. Kahil   {\it{J. Math. Phys.}} {\bf {47}},052501 (2006) \\
{[17]} Magd E. Kahil  {\it{Gravit. Cosmol.}}  {\bf{23}}, 70 (2017) \\
{[18]} Magd E. Kahil {\it{Gravit. Cosmol.}}  {\bf{24}}, 83 (2018) \\
{[19]} Magd E. Kahil {\it{ADAP}} {\bf{3}},  136 (2018) \\

\end{document}